# Deep potential for interaction between hydrated $Cs^+$ and graphene

Yangjun Qin[1,2], Liuhua Mu[3,4], Xiao Wan[1], Zhicheng Zong[1,2], Tianhao Li[1,2], Haisheng Fang[1], Nuo Yang[2,†]

1) School of Energy and Power Engineering, Huazhong University of Science and Technology, Wuhan 430074, China

2) Department of Physics, National University of Defense Technology, Changsha 410073, China

3) Wenzhou Institute, University of Chinese Academy of Sciences, Wenzhou 325001, China.

4) School of Physical Science, University of Chinese Academy of Sciences, Beijing 100049, China

†Corresponding E-mail: N.Y. (nuo@nudt.edu.cn)

## ABSTRACT

The influence of hydrated cation-π interaction forces on the adsorption and filtration capabilities of graphene-based membrane materials is significant. However, the lack of interaction potential between hydrated $Cs^+$ and graphene limits the scope of adsorption studies. Here, it is developed that a deep neural network potential function model to predict the interaction force between hydrated $Cs^+$ and graphene. The deep potential has DFT-level accuracy, enabling accurate property prediction. This deep potential is employed to investigate the properties of the graphene surface solution, including the vibrational power spectrum of water density distribution, radial distribution function, and mean square displacement. Furthermore, the adsorption energy and the amount of charge transferred between hydrated $Cs^+$ and graphene were calculated for varying amounts of bound water, indicating that the presence of water molecules inhibits charge transfer and weakens the interaction between the ions and graphene. The method provides a powerful tool to study the adsorption behavior of hydrated cations on graphene surfaces and offers a new solution for handling radionuclides.

## Introduction

Currently, the global marine ecosystem is facing a significant challenge for radioactive pollution that needs to be addressed[1, 2]. The occurrence of accidents at nuclear facilities and the improper disposal of radioactive waste presents a direct threat to marine organisms and human health[3, 4]. The decontamination of marine environments, particularly the extraction of radioactive pollutants, is a critical yet complex endeavor that is vital for the preservation of the ecological balance and the public health security.

The treatment of radioactive elements in seawater has rapidly developed over the years. Amongst the numerous methodologies employed, adsorption[5] and membrane filtration[3, 6] techniques stand out for their efficiency and ease of operation. These processes serve as potent tools in the arsenal against marine nuclear effluents, finding broad application across diverse sectors.

Adsorption and membrane filtration are highly effective methods for removing radioactive materials and are extensively applied in various fields. Molecular sieves[7], zeolites[8], ion-exchange resins[9], and metal-organic frameworks[10] have emerged as pivotal materials in the efficient sequestration of radiocerium and strontium ions, employing either physical adsorption or chemical bonds strategies. Furthermore, Carbon nanomaterials, when surface-modified[11, 12], enhance ion-material interactions, achieving superior adsorption efficacy. However, the associated costs and operational complexities remain significant hurdles. Graphene, with its unparalleled surface area and cost-effectiveness, emerges as a promising candidate for ion adsorption membranes[13, 14].

Graphene adsorption research has significantly increased in the past two decades. The strong interaction force between cation and graphene has been gradually recognized[15], despite its notable divergence from the prevailing perception. Molecular dynamics[16, 17], a powerful analytical tool, has been instrumental in elucidating hydrated cation-π interactions[18, 19]. Yet, the absence of a precise potential function to describe these interactions has hindered progress. Fang et al.'s pioneering work, combining quantum mechanics and molecular dynamics, led to the successful calibration of the $Na^+$-$\pi$ interaction potential[20, 21]. Their model predicted ion accumulation on graphene surfaces, validating the concept of ion sieving[22]. Other approaches include incorporating cation polarization effects into the empirical potential function[23] and modifying the depth of the potential well for C-Na interactions[24]. Nevertheless, these methodologies have not yet attained the desired level of accuracy, and their practical viability remains to be further discussed and validated.

The interaction potential between hydrated cation is severely limited. In previous, a multitude of quantum mechanical methodologies were employed, encompassing multi-parameter fitting and computations of multi-body interactions, among other sophisticated techniques. However, this approach has been beset with challenges, particularly when applied to the training of potential functions for systems comprising four or more distinct elements, such as hydrated $Cs^+$ systems. The intricacies and computational demands increase exponentially, rendering the traditional quantum mechanical methods less feasible for such complex systems. Recently, machine learning methods[25, 26] offer a powerful tool with the computational efficiency of molecular dynamics and the precision of density functional theory (DFT)[27]. Deep potential (DP)[28-31] stands out for its scalability and accuracy in handling large atomic systems than others[32-35]. DP has been employed in diverse fields[36-39]. The adsorption mechanism of radioactive ions on graphene surfaces could be elucidated by the study of DP potentials containing hydrated $Cs^{+}$-$\pi$ interactions, with implications for both fundamental and applied nuclear wastewater treatment research.

This study constructed a potential for hydrated $Cs^{+}$-$\pi$ interaction using the machine learning potential function method. Firstly, the accuracy of the potential function was validated. Secondly, the properties of the graphene surface solution, including the vibrational power spectrum of the water, the density, the radial distribution function (RDF), and the mean square displacement (MSD) were calculated. Finally, the transfer of charge and energy bands of system was calculated to elucidate the nature of the hydration $cs^{+}$-$\pi$ interaction.

## Deep potential

In the DP model, meticulous parameter tuning is paramount to ensure the precision and reliability of the resultant potential function. The DP framework assumed that the potential energy $E$ of any given configuration can be decomposed into the summation of individual atomic contributions $E_i$, each of which is a function of the local environment descriptor $D_i$ of atom $i$. This descriptor describes the local environment of atom i within the truncation radius. The truncation radius and smoothing radius are set to 0.6 nm and 0.05 nm, respectively, striking a balance between computational efficiency and the accuracy of the interatomic potential. The dimensions of the embedding network and the fitted network are (25, 50, 100) and (240, 240, 240), respectively. Furthermore, the hyper-parameters start-pref_e, start-pref_f, start-pref_v, limit-pref_e, limit-pref_f, and limit-pref_v, which regulate the weights of energy and force losses in the total loss function, have been set to 0.02, 1000, 0, 1.0, 1.0, and 0, respectively. The initial learning rate is set at $10^{-3}$ and decays exponentially to $10^{-8}$ at the conclusion of the training period. The number of training steps is set to 1,500,000. The functions trained with these optimized parameters have been demonstrated to achieve the training accuracy of DFT[39, 40]. thereby substantiating the rationality and effectiveness of the adopted parameterization strategy.

The initial data obtained in ab initio molecular dynamics (AIMD) is insufficient to encompass the entirety of the structural phase space. The DP generator package[41, 42] is utilized to orchestrate the Large-scale Atomic/Molecular Massively Parallel Simulator software. This enables an exhaustive exploration of the structural phase space and facilitates the procurement of structurally valid data. So, four models were crudely trained utilizing the preliminary dataset. Subsequently, MD simulations were conducted across the temperature range utilizing a DP model. Within these simulations, a comparative analysis was performed, focusing on the fluctuation of energies and atomic forces exerted by structures at distinct time points. Additionally, one DP model was leveraged to compute predictions for the remaining three potential function models, fostering a comprehensive comparison and validation process. The maximum force deviation $\delta_f^{max}$ equations for the four models are as follows:

$$\delta_f^{max} = max_i\sqrt{\langle|F_i - \langle F_i\rangle|^2\rangle} \tag{1}$$

When the atomic force $\delta_f^{max}$ is smaller than $\delta_{low}$, the configuration is labeled as exact configuration, while when $\delta_f^{max}$ is larger than $\delta_{high}$, the configuration is labeled as failed configuration. When $\delta_{low}$<$\delta_f^{max}$ <$\delta_{high}$, the configuration is labeled as a candidate configuration, which will be added to the initial data set for training in the next step. A total of 15 iterations were performed throughout the simulation, as detailed in Table S1.

AIMD is performed based on Vienna ab initio simulation package with the Perdew–Burke–Ernzerhof generalized gradient approximation and the projector augmented wave pseudopotentials. The base training dataset was extracted from energy, force and virial data obtained from an ensemble of 2,000 initial configurations. These configurations were subjected to a series of 10-step AIMD simulations. The temperature parameters for these simulations ranged from 200 K to 300 K, and the simulation conditions were 0 Pa with timestep of 0.5 fs[43]. As illustrated in Fig. S1, our study definitively demonstrates that an energy cutoff threshold of 520 eV in conjunction with a k-point spacing of 0.3 Å$^{-1}$ is sufficient to achieve convergence of energy values and atomic forces. The k-point grid has been designed with a size of $3 \times 3 \times 1$, ensuring a balanced trade-off between accuracy and computational efficiency.

The DP-gen was employed to train the interaction potential between hydrated $Cs^+$ and graphene. As previously outlined, the training process encompasses three key components within the potential function. This iterative training process continues until a precision benchmark of 99% is achieved, at which point the training is terminated.

The accuracy of the DP potential function is corroborated by a comparison of the results with the DFT calculations, as depicted in Fig. 1(a) through 1(d). The data set for training and validation purposes comprises 7,244 structures and 1,000 structures, respectively. The root mean square error (RMSE) values, which serve as measures of accuracy, are reported to be 4.25 meV/atom for the training dataset and 2.58 meV/atom for the validation dataset, as illustrated in Fig. 1(a) and 1(c). Further analysis in Fig. 1(f) elucidates the potential function curves for hydrated cation structures, revealing the concordance between the DFT and DP, as well as illustrating that the universal force field (UFF)[44] differs significantly from the DP . The high degree of accuracy exhibited by DP renders them eminently suitable for deployment in molecular dynamics simulations, ensuring reliable predictions and insights into complex systems.

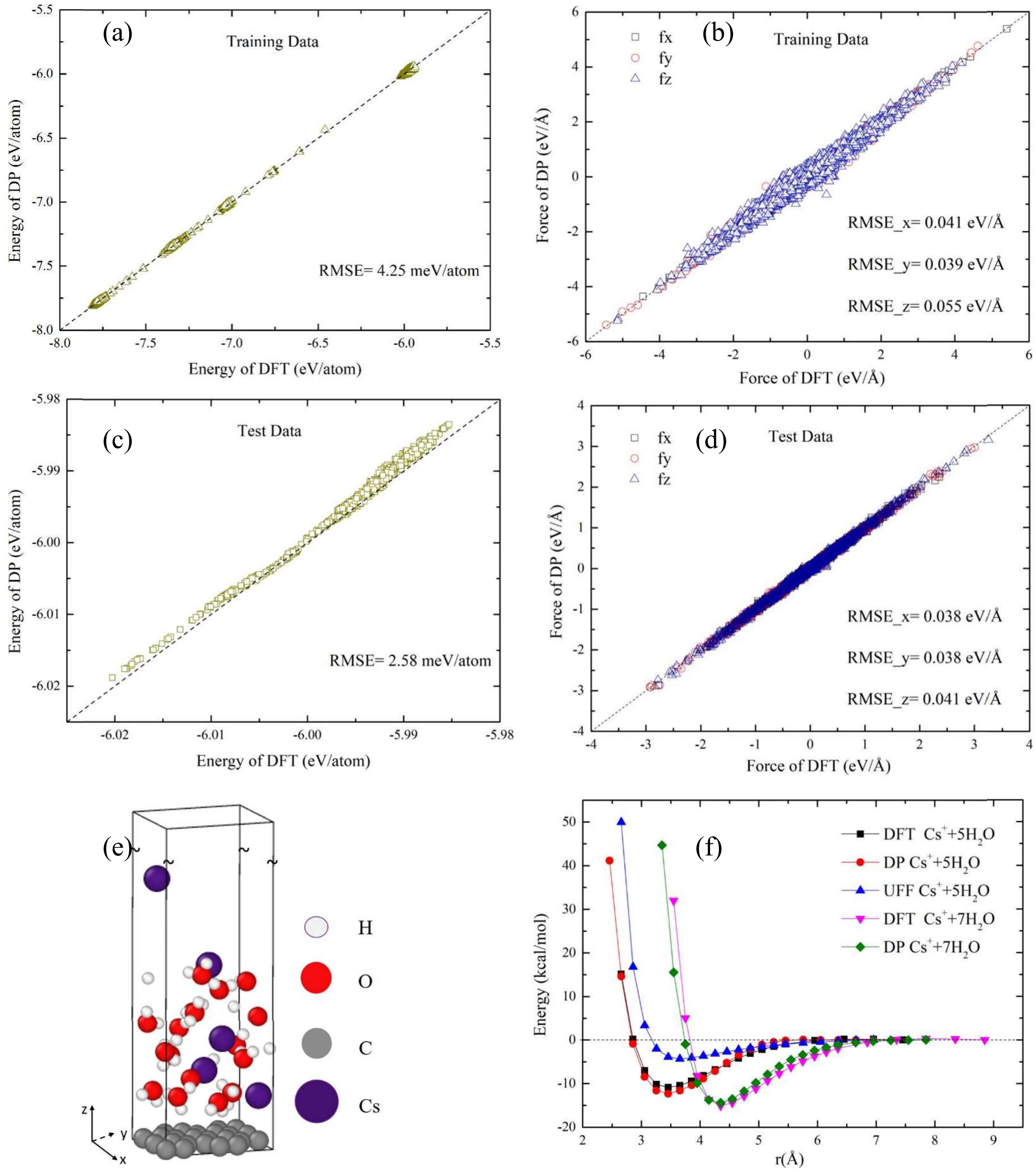

Figure 1. The training and validation results of DP. (a) The system energy by DP (the training set) versus the energy by DFT. (b) The force for the training set in comparison with the results of the DFT calculations. (c) Energy for the test set versus the results of the DFT calculations. (d) Force for the training set versus the results of the DFT calculations. (e) Schematic structure of the system, including $H_2O$, $Cs^+$ and graphene. (f) The comparisons of potential function between the hydrated $Cs^+$ and the graphene, calculated by DFT, DP and UFF.

## Simulation results and discussions

The objective of this investigation is to examine the impact of graphene π-bonds on the characteristics of its surface solution and to assess the discrepancies between the DP and UFF for property prediction. Two distinct systems, comprising 3040 atoms, were constructed using molecular dynamics simulations as visualized with OVITO[45] in Fig. 2. A significant distinction between these two systems is that classical molecular dynamics is unable to utilize the UFF for calculations in non-neutral systems, whereas DP does not incorporate the impact of charge in molecular dynamics simulations. Detailed UFF parameters are referenced in Ref.[20],

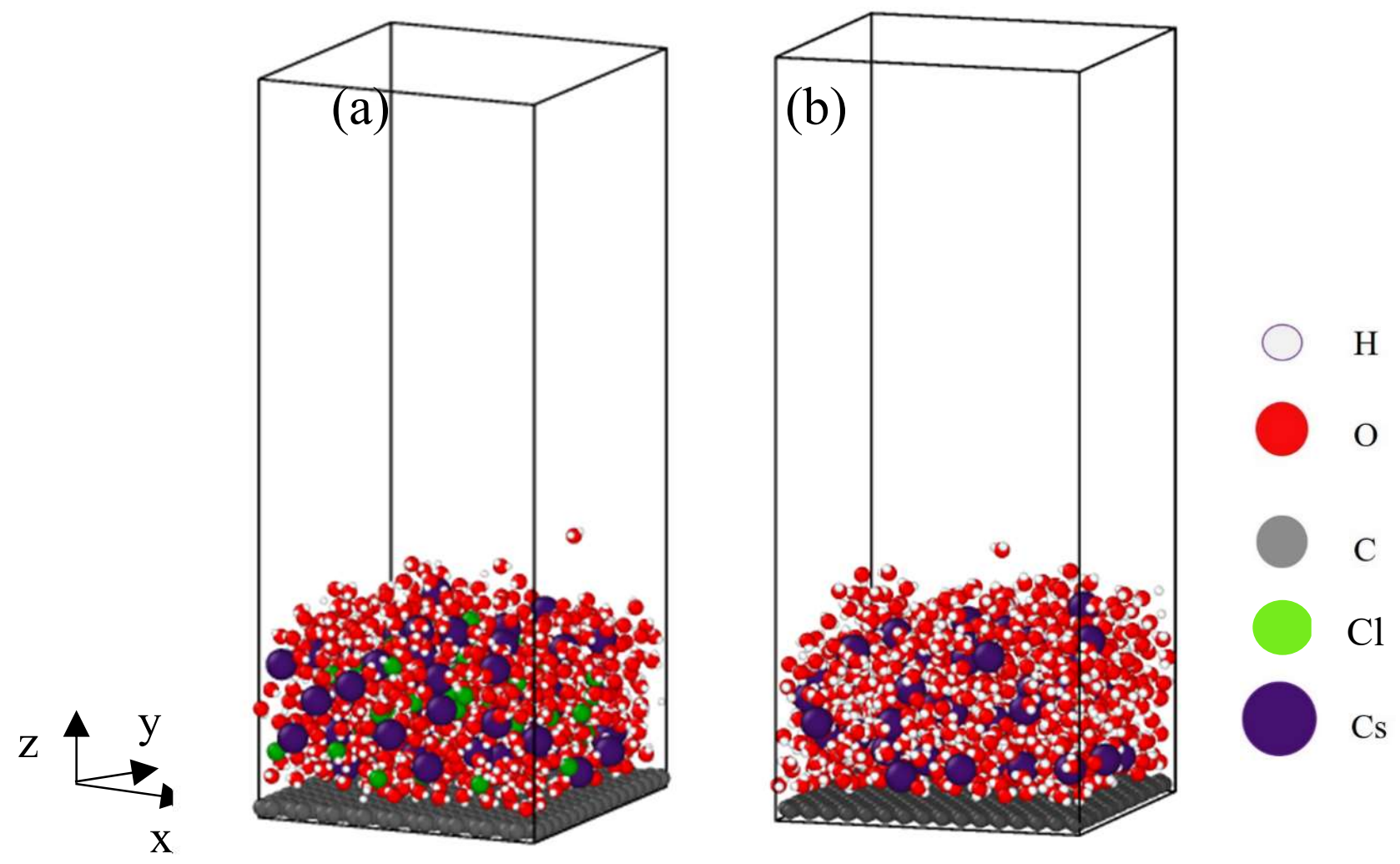

Figure 2. The snapshots of the system. (a) the system of UFF, (b) the system of DP.

The vibrational power spectrum of water is an important parameter of aqueous solutions. Furthermore, this section examines the impact of graphene π-bonds on the vibrational power spectrum of aqueous solutions, which facilitates an understanding of molecular vibrations. The vibrational power spectrum of a water molecule is determined through Fourier transformation of the autocorrelation function of the velocity of water. The latter can be expressed as [46]:

$$I(\nu) \propto \lim_{\tau\to\infty} \int_{-\tau}^{\tau} \langle v_i(t_0 + t) \cdot v_j(t_0) \rangle e^{-i2\pi\nu t} dt \tag{2}$$

where $I$ is the intensity, $\nu$ is the vibrational frequency, and $v$ is the velocity of the atom. After counting the data for another 50 ps and calculating the vibrational power spectrum of water containing two regions: the range of 2 nm to 4 nm (region 1) and the range of 0.6 nm from the graphene wall (region 2).

It is investigated that the effect of π-bonds in graphene on the vibrational power spectrum of water molecules. As illustrated in Fig. 3(a), the velocity spectrum reveals a multiplicity of distinct peaks, each attributable to specific interactions: hydrogen bonding, the H-O-H bend angle, and O-H bond formation, respectively. A comparison of the theoretical and experimental data reveals a clear agreement between the peak frequencies of the water vibration power spectrum in region 1 and the corresponding experimental[47, 48]. In the high-frequency range, the vibrational power spectrum of water in region 2, in proximity to graphene, is observed to be redshifted. The vibrational power spectrum of water molecules, as portrayed in Fig.S2, was computed under the application of UFF. It has been revealed through the investigations that there exists a considerable disparity between these computations and experimental outcomes, thus indicating an insufficiency in accurately delineating the inherent properties of water through this methodology.

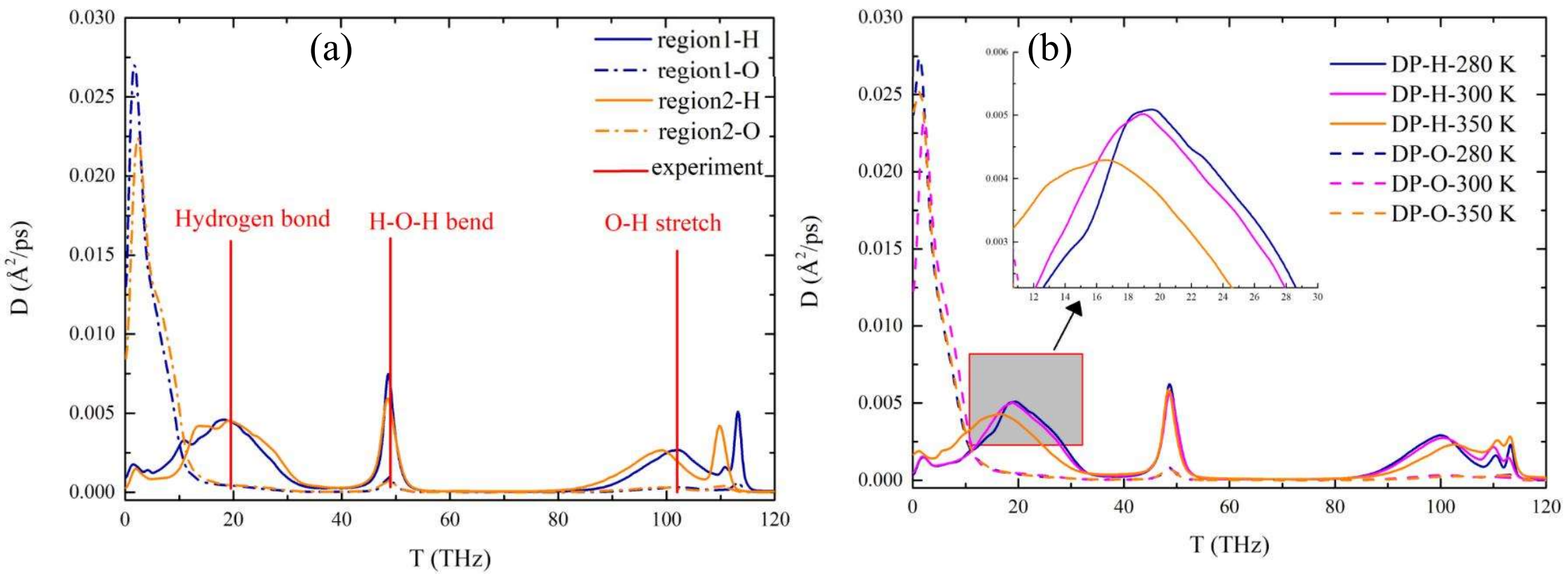

Figure 3. (a) The comparison of power spectra of water calculated by DP and experiment[47] at 280 K. (b) The power spectra of water at three different temperatures calculated by DP.

The effect of temperature on the vibrational power spectrum of water was also investigated. It was found that in Fig. 3(b), the low-frequency region of the vibrational power spectrum undergoes a redshift as temperature ascends. This spectral shift is predominantly rooted in the thermal-induced restructuring of the hydrogen bonding network, a dynamic interplay that subsequently modulates the intermolecular interactions within the network. At elevated temperatures, the stability of the hydrogen bonding network is compromised, leading to a diminution in the strength of hydrogen bonding interactions. This thermally-induced weakening facilitates the disruption of the hydrogen bonding network, thus impacting the vibration frequency. In Fig. 3(b), the temperature has a significant effect on the vibrational absorption peaks of the hydrogen bonding network of water. However, it has a limited effect on the absorption peaks of the bending and stretching vibrations of water. In conclusion, the presence of π-bonds in graphene affects the motion of water molecules within the ionic solution, which in turn affects the properties.

The density profile is a key parameter that assess the influence of the cation-π in the system. In this segment, the density of the system under 280 K,300 K and 350 K is investigated utilizing the LAMMPS with DP and UFF potentials. As illustrated in Fig. 4(a), the results reveal that the density of graphene surface as calculated using the DP is approximately triple that determined through the UFF. This pronounced disparity underscores the superior strength of the interaction forces exerted on graphene when employing the DP model.

Furthermore, a phenomenon was observed with respect to the thickness of the high-density layer near the graphene surface. It was found that the thickness of the liquid film increased from 2 to 3 nm at elevated temperatures. For the DP, a separation of molecular clusters occurred at 350 K. This phenomenon can be attributed to the increase in the kinetic energy of the system ions, which enhances the intermolecular motion.

Then, the RDF for oxygen-oxygen (O-O), oxygen-hydrogen (O-H), and hydrogen-hydrogen (H-H) pairs were calculated using DP and UFF. These results were also compared with previous studies on pure water, as shown in Fig. 4(b).

For the O-O interactions, a prominent peak is observed at approximately 2.75 Å in systems, indicating a clear short-range order within the system. The intensity of this peak varies among the systems, reflecting differences in the probability distribution of oxygen atoms at this distance. Specifically, the DP model exhibits a higher peak intensity compared to the UFF model, suggesting stronger and more localized clustering of oxygen atoms. For the O-H interactions, each model reveals a main peak at around 0.95 Å, which is consistent with the typical length of hydrogen bonds.

Regarding the H-H interactions, the first significant peak appears at approximately 1.6 Å in all models, representing the closest approach between hydrogen atoms.

In summary, the DP model consistently predicts stronger and more localized interactions compared to the UFF model, with higher precision. the fundamental divergence in interaction forces, as encapsulated by the DP and UFF models, stands as the cardinal driver behind the enrichment effect observed on the graphene surface. This effect, in turn, cascades into discernible alterations in surface density distribution and intrinsic wettability characteristics[49]. These deviations from the predictions rendered by the UFF method underscore the necessity for adopting advanced computational models, such as the DP framework, to accurately decipher and represent the complex interplay between hydrated cation and graphene surfaces. These insights differ from the prevailing view and highlight the key role that deep learning plays in elucidating the properties of ionic solutions on graphene surfaces.

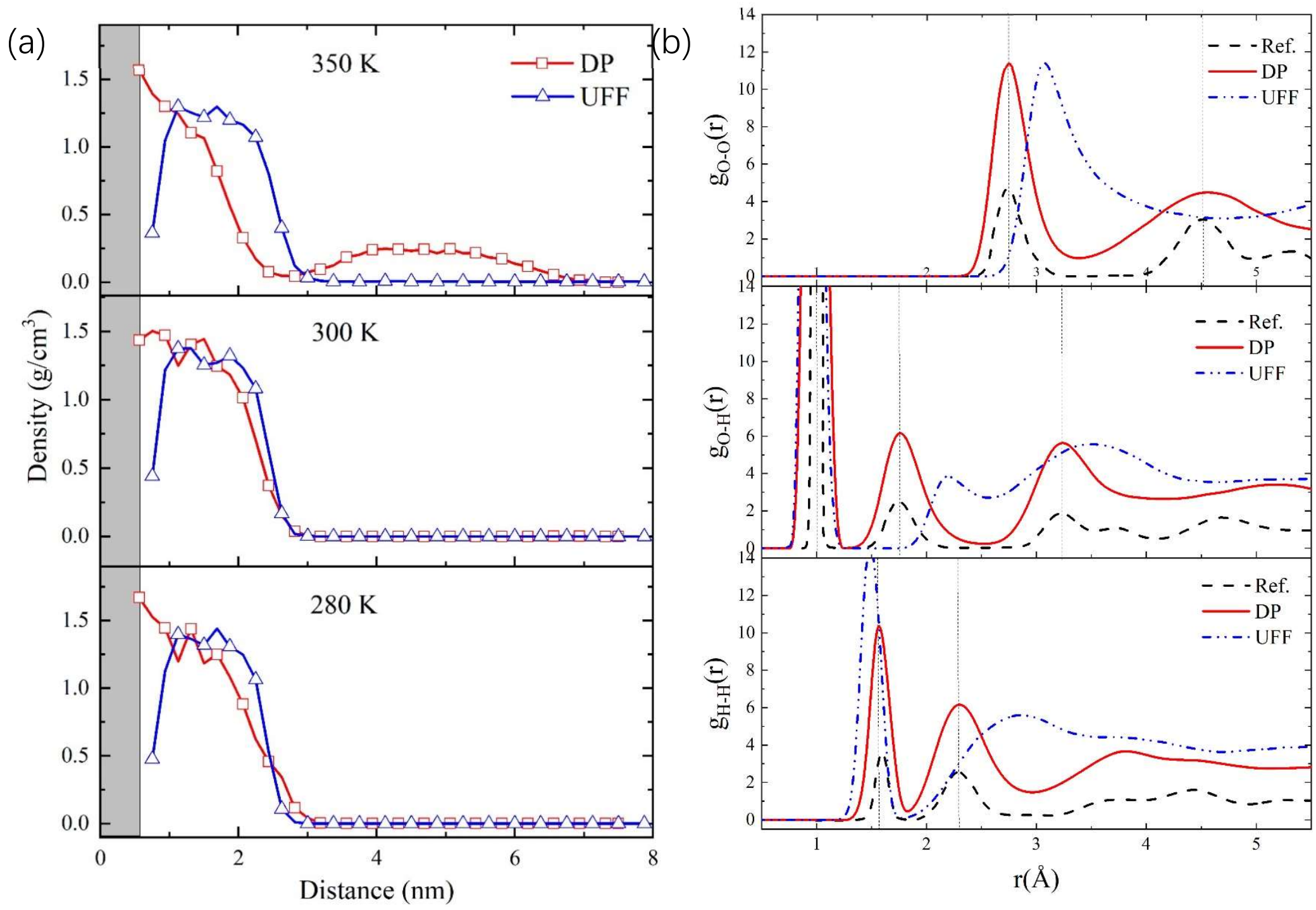

Figure 4. (a)Density profile of the solution under different temperature (the grey areas represent single-layer graphene wall). (b) Comparison of RDF of O-O,O-H, and H-H in DP,UFF ,and Ref.[29] at 280 K.

The MSD is a covariate that quantifies the diffusion capacity, facilitating a nuanced exploration of the impact exerted by hydrated cation-π interactions on diffusive processes. MSD distribution in the z-direction (orthogonal to the graphene plane) for the ion solution, meticulously computed at 280 K, was calculated quantitatively in Fig. 5(a). A substantial disparity exists between the MSD values derived from the DP and UFF potential. The UFF potential demonstrates a swifter attainment of equilibrium, accompanied by a more pronounced fluctuation in MSD values as temperatures rise. Furthermore, the MSD of $Cs^+$ and water in solution was calculated in Fig.5(b). A notable discrepancy was observed between the MSD distributions calculated by the two potential functions. At 1 ns, the MSD of water calculated with the DP potential was four times higher than the value calculated by UFF, and two times higher for $Cs^+$. These findings unquestionably demonstrate the existence of unique interaction forces between hydrated cation and graphene (cation-π interaction), which are not accounted for by the universal force field.

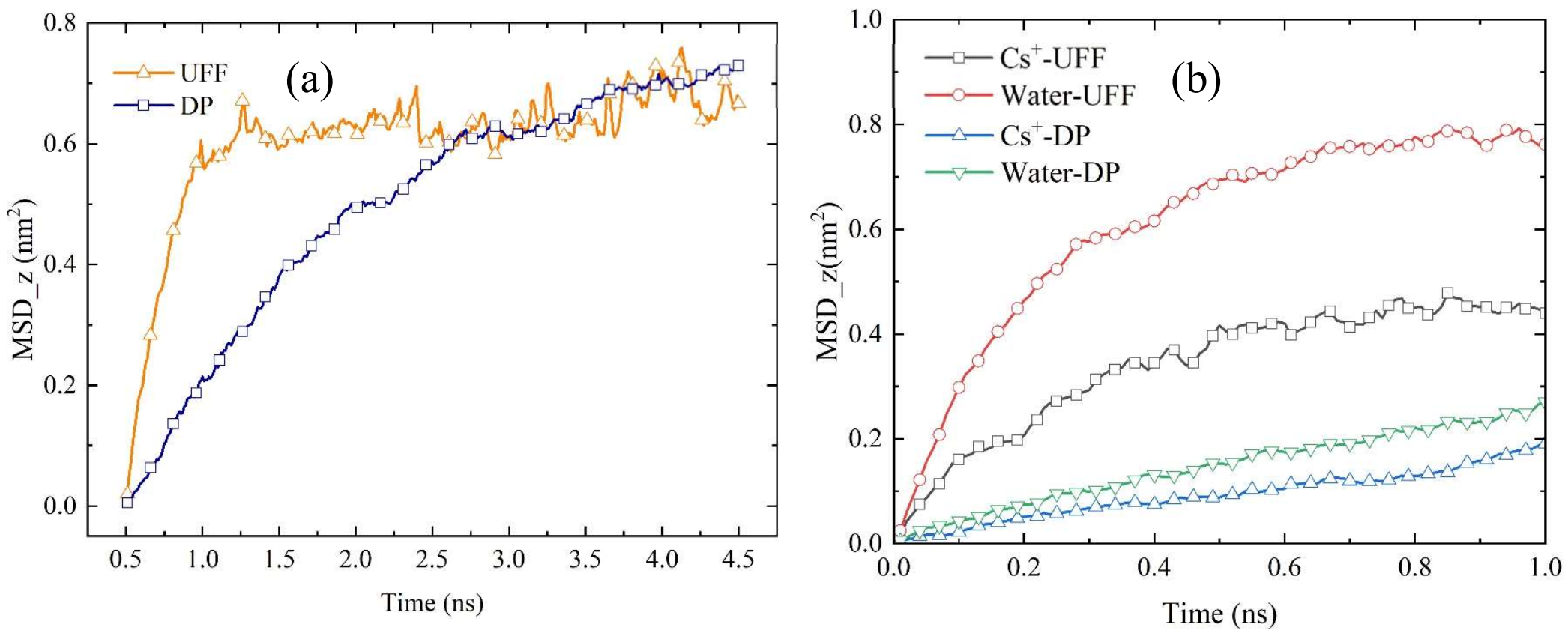

Figure 5. (a) The comparisons of MSD of the solution along z-direction by using UFF and DP at 280 K. (b) MSD distribution of water molecules and $Cs^+$ in solution.

To understand the adsorption mechanism of hydrated $Cs^+$ on graphene surfaces, the energy bands were calculated. As depicted in Fig.6, it shows the energy bands for both pristine graphene and graphene with adsorbed $Cs^+$. Fig.6(a) illustrates the pristine graphene's energy band structure, characterized by the intersection of valence and conduction bands at the distinctive Dirac point, indicative of the material's unique electronic properties. Upon the adsorption of $Cs^+$ onto the graphene surface, as shown in Fig. 6(b), a notable splitting of energy levels occurs around the Fermi energy, leading to an enhanced density of electronic states. The DOS distribution, presented in Fig. S3, exhibits a pronounced enhancement within the dashed box region, signifying an increased electron density. This observation points towards a robust interaction between graphene and hydrated $Cs^+$, attributed to the ion-π interaction.

In aqueous environments, water molecules play a crucial role in ion-π interactions. As shown in Fig. 6(c), an increase in the number of bound water molecules leads to a weakening of the $Cs^+$-π interaction and an increase in the equilibrium distance. The most stable adsorption configuration is illustrated in Fig. S4. The number of C atoms in the system is 180, as Fig. S5 provides sufficient evidence to confirm the accuracy of the adsorption energy calculations.

To investigate this phenomenon, born effective charge[50, 51] was calculated using VASP. The results showed that the charge transferred from graphene to the hydrated ion molecular orbitals decreases. This finding provides a clear explanation for the weakening of the $Cs^+$-$\pi$ interaction.

On the other hand, calculations of the energy bands of the system reveal that an increase in the number of water molecules results in new electronic energy levels and significant flat bands in the electronic energy bands of graphene. This indicates a significant decrease in electron mobility, inhibiting charge migration between graphene and hydrated ions. Thus, the presence of water molecules shields the interaction between $Cs^+$ and graphene, leading to the trend: graphene- $Cs^+$ > graphene- $Cs^+$-$5H_2O$ > graphene- $Cs^+$-$7H_2O$.

Our findings offer novel insights into the adsorption behavior of hydrated ions, diverging from previous observations[52]. For instance, the shift in the vibrational frequency of the O-H bond, as discussed in Fig. 3(a), can be reasonably attributed to the redistribution of electrons among the orbitals of graphene and water molecules.

In this discussion, we have conclusively elucidated that the interaction force between hydrated cations and the graphene surface is fundamentally due to the intricate interplay of electron migration. Specifically, this phenomenon involves electron transfer between graphene's intrinsic $\pi$-electron cloud and the hydrated ions. This electrostatic force arises from the redistribution of charge density. This results in a strong electrostatic force that plays a key role in consolidating the affinity of hydrated cations for graphene. In essence, our findings not only elucidate the microscopic mechanism of the interaction between hydrated cations and graphene, but also contribute to a broader understanding of the interaction between ions and π-electrons. This knowledge provides a theoretical foundation for subsequent ion adsorption and ion filtration.

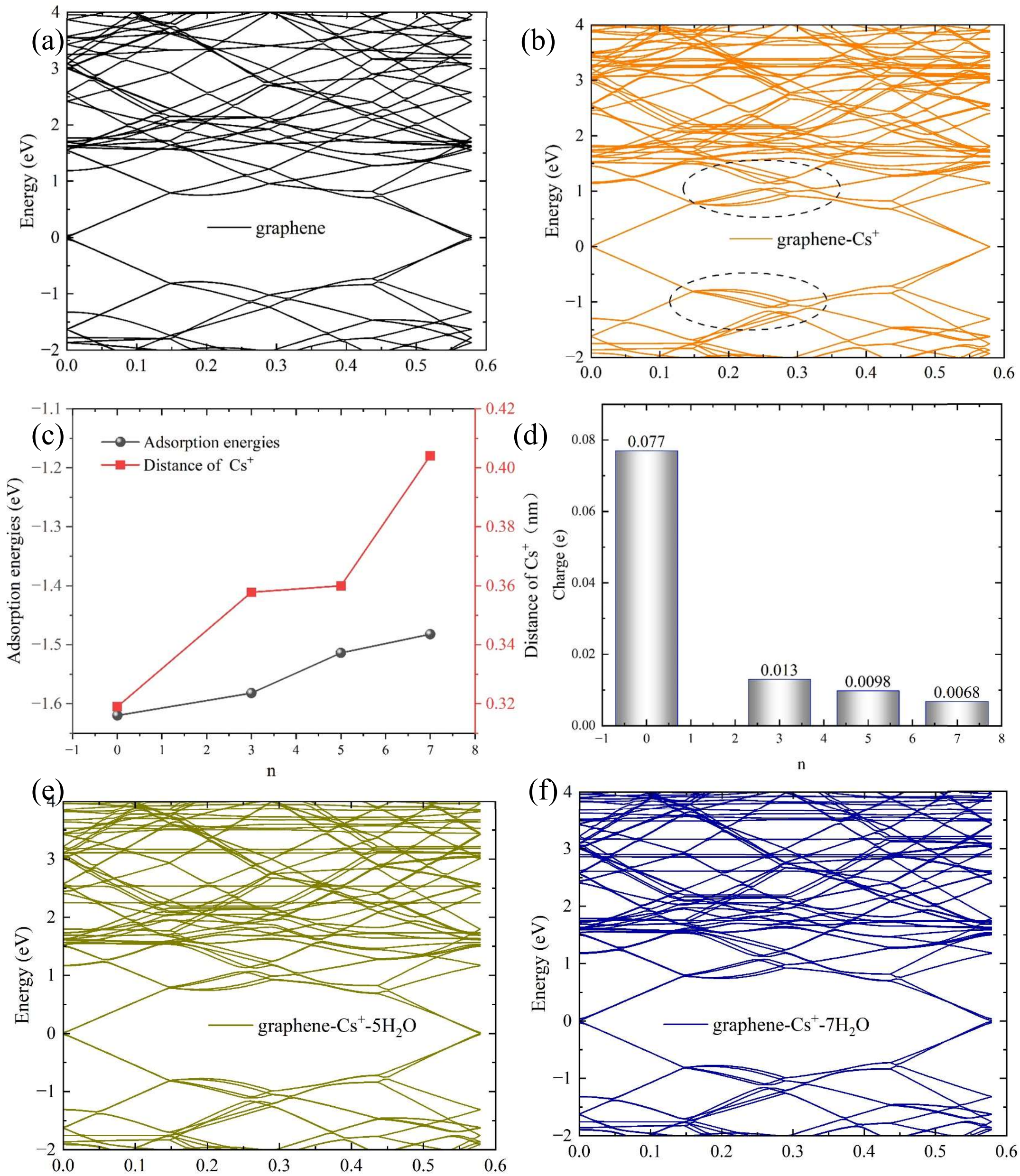

Figure 6. (a) Band structure of graphene. (b) Band structure of graphene-$Cs^+$. (c)The adsorption

energy and equilibrium distance between graphene and hydrated $Cs^+$. (d) Charge transfer between graphene and $Cs^+$. (e) Band structure of graphene-$Cs^+$-$5H_2O$. (f) Band structure of graphene-$Cs^+$-$7H_2O$

## Conclusion

Here, machine learning techniques were employed to train and validate the hydrated $Cs^+$-π interaction deep potential and predict solution properties. The accuracy of DP is much higher than UFF by comparisons.

By using molecular dynamics simulation, it is found that the water vibration power spectrum calculated by DP matches well with the experimental results. And the presence of π-bonds dampens the vibration frequency at 102 THz. This is mainly due to electron transfer affecting the strength of the hydroxyl bonds.

Furthermore, the distribution of density on the surface of graphene was calculated. It was found that the density calculated by DP was three times higher than that of UFF. The MSD along the z-direction, calculated by DP, was found to be significantly smaller than that calculated by UFF. This can be attributed to the strong interaction forces between graphene and $Cs^+$, which cannot be described by the UFF.

The hydrated $Cs^+$-π interaction force is attributed to the orbital electrons of $Cs^+$, which cause a cleavage of the energy bands and an increase in the electronic states of the system. Simultaneously, the presence of water molecules leads to flatter bands in the graphene system and reduces charge migration.

These findings not only deepen the understanding of the intricate interactions between hydrated ions and the graphene surface, but also have important implications for the adsorption and removal of radioactive ions on graphene substrates. A theoretical framework is established for the development of graphene-based materials with enhanced ion adsorption and filtration capabilities.

## Conflicts of interest

**Authorship contribution statement**: Yangjun Qin: Investigation, Writing - original draft, Data curation, Formal analysis. Liuhua Mu: Investigation, Writing - original draft, Software. Xiao Wan: Investigation. Zhicheng Zong: Investigation, Software. Tianhao Li: Investigation, Software. Haisheng Fang: Investigation. Nuo Yang: Project administration, Conceptualization, Writing - review & editing.

## Acknowledgements

This work was sponsored by the National Key Research and Development Project of China, no. 2018YFE0127800. The work was carried out at National Supercomputer Center in Tianjin, and the calculations were performed on Tianhe new generation supercomputer